\documentclass[manuscript]{aastex}

\def\kms{\hbox{km s$^{-1}$}}
\def\VLSR{\hbox{$V_{\rm LSR}$}}

\def\sun{\hbox{$\odot$}}
\def\earth{\hbox{$\oplus$}}
\def\lesssim{\mathrel{\hbox{\rlap{\hbox{\lower4pt\hbox{$\sim$}}}\hbox{$<$}}}}
\def\gtrsim{\mathrel{\hbox{\rlap{\hbox{\lower4pt\hbox{$\sim$}}}\hbox{$>$}}}}

\def\arcdeg{\hbox{$^\circ$}}
\def\arcmin{\hbox{$^\prime$}}
\def\arcsec{\hbox{$^{\prime\prime}$}}
\def\dotdeg{\hbox{$.\!\!^\circ$}}
\def\dotsec{\hbox{$.\!\!^{\prime\prime}$}}

\slugcomment{To appear in the Astrophysical Journal Letters}
\shorttitle{Signature of an Intermediate-Mass Black Hole}
\shortauthors{Oka et al.}

\begin{document}

\title{Signature of an Intermediate-Mass Black Hole in the Central Molecular Zone of Our Galaxy}

\author{Tomoharu Oka$^{1, 2}$, Reiko Mizuno$^1$, Kodai Miura$^2$, \&\ Shunya Takekawa$^2$}
\affil{$^1$Department of Physics, Institute of Science and Technology, Keio University, 3-14-1 Hiyoshi, Yokohama, Kanagawa 223-8522, Japan}
\affil{$^2$School of Fundamental Science and Technology, Graduate School of Science and Technology, Keio University, 3-14-1 Hiyoshi, Yokohama, Kanagawa 223-8522, Japan}
\email{tomo@phys.keio.ac.jp}

\begin{abstract}
We mapped the high-velocity compact cloud CO--0.40--0.22 in 21 molecular lines in the 3 mm band using the Nobeyama Radio Observatory 45 m radio telescope. Eighteen lines were detected from CO--0.40--0.22.  The map of each detected line shows that this cloud has a compact appearance ($d\!\simeq\!3$ pc) and extremely broad velocity width ($\Delta V\!\simeq\!100$ \kms).  The mass and kinetic energy of CO--0.40--0.22 are estimated to be $10^{3.6}$ $M_{\sun}$ and $10^{49.7}$ erg, respectively.  The representative position--velocity map along the major axis shows that CO--0.40--0.22 consists of an intense region with a shallow velocity gradient and a less intense high-velocity wing.  Here, we show that this kinematical structure can be attributed to a gravitational kick to the molecular cloud caused by an invisible compact object with a mass of $\sim\!10^5$ $M_{\sun}$.  Its compactness and the absence of counterparts at other wavelengths suggest that this massive object is an intermediate-mass black hole.
\end{abstract}

\section{Introduction}
Most galaxies, including the Milky Way, are thought to have black holes (BHs) with masses greater than a million solar masses ($M_{\sun}$) at their centers.  However, the origins of such supermassive black holes (SMBHs) remain unknown (e.g., Djorgovski et al. 2008).  One possible scenario is that a ``seed'' BH with tens or hundreds of solar masses at the center of a galaxy grows by accretion of matter (e.g., Volonteri \&\ Rees 2005).  Another scenario is that black holes with masses of $\sim\!10^3\,M_{\sun}$, which are formed by runaway coalescence of stars in young compact star clusters (Portegies Zwart et al. 1999), merge at the center of a galaxy to form an SMBH (Ebisuzaki et al. 2001).  The latter theory is slightly more persuasive, as it naturally explains the tight correlation between the mass of a given central SMBH and that of the associated galactic bulge.  To confirm the merging scenario, the ability to unambiguously detect intermediate-mass black holes (IMBHs; $M\!=\!10^{2\mbox{--}5}\, M_{\sun}$) is essential.  Many candidates for IMBHs have been proposed to date on the basis of  their ultraluminous nature (Fabbiano 2006; Roberts 2007), low-temperature blackbody spectral components (Miller et al. 2003), and quasi-periodic oscillations (Casella et al. 2008).  One of the promising IMBH candidate is ESO 243-49 HLX-1, which is the brightest ultra-luminous X-ray source (ULX) outside galactic nuclei (Godet et al. 2009).  However, the nature of ULXs is uncertain, and significant evidence exists that they are not IMBHs (e.g., Ebisawa et al. 2003; Liu \&\ Mirabel 2005; Okajima et al. 2006).  IMBHs have been inferred in several globular clusters from velocity dispersion profiles (Gebhardt et al. 2002; Gerssen et al. 2002; Noyola et al. 2008), but they are often not confirmed by other teams (e.g., Strader et al. 2012).  None of these IMBH candidates are widely accepted as definitive, thus their existence is a long-standing controversy.  In this Letter, an alternative approach to seeking IMBHs based on analysis of the kinematical structures of certain compact clouds is proposed.

The nucleus of the Milky Way Galaxy (Sgr A$^*$) also harbors a 4 million $M_{\sun}$ BH (e.g., Ghez et al. 2008; Gillessen et al. 2009).  A large expanse of dense interstellar gas exists within several hundred parsecs of the nucleus, which is known as the central molecular zone (CMZ; Morris \&\ Serabyn 1996).  While investigating CO {\it J}=3--2 survey data of the Galactic CMZ obtained with the Atacama Submillimeter Telescope Experiment (ASTE) 10 m telescope (Oka et al. 2012), we noticed a peculiar molecular cloud at galactic longitude $-0.40\arcdeg$ and galactic latitude $-0.22\arcdeg$, with local standard of rest (LSR) velocities ranging from $-120$ to $-20$ \kms .  This CO--0.40--0.22 is a compact cloud ($<\!5$ pc at 8.3 kpc distance) with an extremely broad velocity width ($\sim\!100$ \kms ) and a very high CO {\it J}=3--2/{\it J}=1--0 intensity ratio ($\geq\!1.5$; Fig. 1).  The high ratio implies that the cloud consists of dense, warm, and moderate opacity gas (Oka et al. 2007; 2012).  It belongs to a peculiar category of molecular clouds, namely, high-velocity compact clouds (HVCCs)\footnotemark[1], which were originally identified in the CO {\it J}=1--0 emission survey data (Oka et al. 1998).  The position of CO--0.40--0.22 is approximately $0.2\arcdeg$ Galactic southeast of the massive star-forming region Sgr C, being displaced by $\sim\!60$ pc in projected distance from the Galactic nucleus.   Its compact appearance and broad velocity width are remarkable in the HCN {\it J}=4--3 emission image.  It is located at the rim of an expanding shell of dense molecular gas (shell 2; Tanaka et al. 2014) .  Although an H$\alpha$ emission region $0\dotdeg3\!\times\!0\dotdeg3$ in size, RCW137 (Rodgers et al. 1960), overlaps CO--0.40--0.22 in the plane of the sky, it has a positive LSR velocity and is therefore physically unrelated to CO--0.40--0.22.

\footnotetext[1]{The term came from its first example, CO 0.02--0.02, which has a large systemic velocity as well as a large velocity dispersion (Oka et al. 1999).  We identified a number of clouds similar to CO 0.02--0.02 in the CMZ, and named them ``high-velocity compact clouds" since most of them also have large systemic velocities relative to possible parent clouds (Oka et al. 2001; 2012).  Although CO--0.40--0.22 and CO--0.30--0.07 do not have parent clouds and thus their systemic velocities are ambiguous, they share the fundamental properties of HVCCs.}

\begin{figure}[htbp]
\begin{center}
\includegraphics[width=1.0\textwidth]{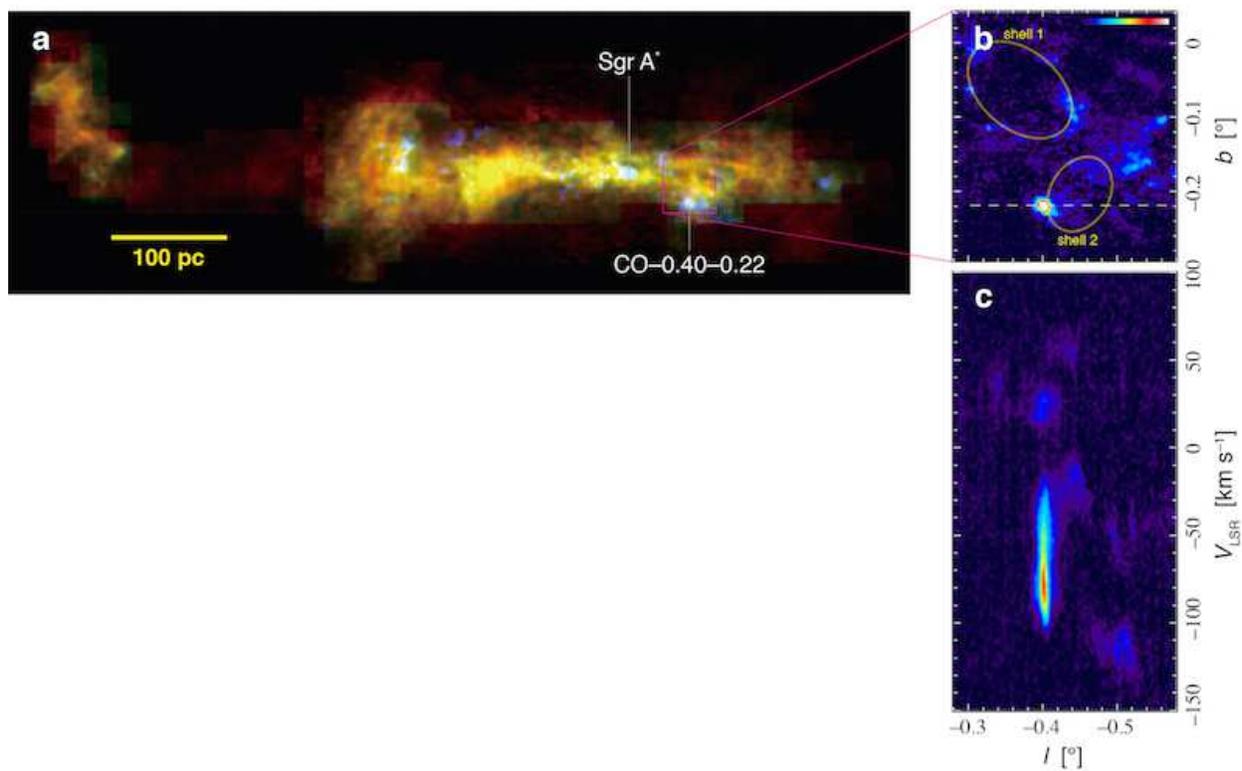}
\end{center}
\caption{(a) Composite CO image of the Galactic CMZ.  Red indicates the velocity-integrated intensity of the CO {\it J} = 1--0 line, yellow indicates that of the CO {\it J} = 3--2 line, and blue is the CO {\it J} = 3--2 emission integrated for data with a high CO {\it J} = 3--2/{\it J} = 1--0 intensity ratio ($\geq\!1.5$).  (b) Magnified map of velocity-integrated HCN {\it J} = 4--3 emission.  The velocity range for integration is $\VLSR\!=-\!120$ to $-20$ \kms .  Loci of the expanding shells are presented by yellow ellipses (Tanala et al. 2014).  (c) Longitude--velocity map of HCN {\it J} = 4--3 line at $b\!=\!-0.22\arcdeg$.  }
\label{fig1}
\end{figure}

\section{Observations}
We performed mapping observations of CO--0.40--0.22 in 21 molecular lines  in the 3 mm band using the Nobeyama Radio Observatory (NRO) 45 m radio telescope.  Table 1 summarizes the frequencies, molecules, transitions, and parameters calculated for each line.  These lines were selected to trace CO--0.40--0.22 on the basis of the result of a 3 mm band line survey toward this cloud with the Mopra 22 m telescope (Oka et al. 2014).  The observations were made 25 January--3 February and  25--28 March 2014.

We used the TZ1 V/H receivers, which were operated in the two-sideband mode.  The system noise temperatures ranged from 120 to 360 K at $EL\!=\!70\arcdeg$ during the observations.  We observed the $3\arcmin\!\times\!3\arcmin$ area around CO--0.40--0.22 in the on-the-fly mapping mode.  All data were obtained by position switching between target scans and the clean reference position, $(l, b)\!=\!(0\arcdeg, -1\arcdeg)$.  The antenna temperature was calibrated by the standard chopper-wheel method.  We used the SAM45 spectrometers in the 1 GHz bandwidth (244 kHz resolution) mode.  The frequency resolution corresponds to a 0.8 \kms\ velocity resolution at 86 GHz.  The half-power beamwidth of the NRO 45 m telescope is $\simeq\!20\arcsec$ at 86 GHz.  Pointing errors were corrected every 2 h by observing the SiO maser source VX Sgr at 43 GHz.  The pointing accuracy of the telescope was good to $\leq\!3\arcsec$  in both azimuth and elevation.

The obtained data were reduced using the NOSTAR reduction package.  We subtracted the baselines of all spectra by fitting first- or third-order polynomial lines.  We scaled the antenna temperature by multiplying it by $1/\eta_{\rm MB}$ to obtain the main-beam temperature, $T_{\rm MB}$.  The main-beam efficiencies of the TZ1 V/H receivers at the line frequencies were calculated as $\eta_{\rm MB}(\nu)\!= a\,{\rm exp}\left(-(\nu/b)^2 \right)$, where $a$ and $b$ were obtained by least-square fitting to those measured at three frequencies (86, 110, and 115 GHz).  All the data were resampled onto a $7\dotsec5\!\times\!7\dotsec5\!\times\!2$ \kms\ grid to obtain the final maps.

\begin{deluxetable}{lcccccccc}
\tabletypesize{\scriptsize}
\tablecaption{Observed Line Parameters.\label{tbl-1}}
\tablewidth{0pt}
\tablehead{
\colhead{Freq.} & \colhead{Molecule} & \colhead{Transition} & \colhead{$T_{\rm MB}$\tablenotemark{a}} & \colhead{$S$\tablenotemark{b}}  & \colhead{$\sigma_{\rm V}$} & \colhead{$e$} & \colhead{P.A. } \\
 \colhead{[GHz]}  &  &  & \colhead{[K]}  & \colhead{[pc]}  & \colhead{\scriptsize{[\kms ]}}  &    &  \colhead{[\arcdeg]}
}
\startdata
81.881  &  HC$_3$N & {\it J}=9--8 &  $2.03\pm 0.10$  &  1.17  &  21.6  &  0.66  &  42.9 \\
85.339 &   c-C$_3$H$_2$ & $2_{12}$--$1_{01}$ &	 $0.30\pm 0.06$  &  1.19  &  29.4  &  0.33  &  $-42.5$ \\
86.340  &  H$^{13}$CN & {\it J}=1--0  & $1.86\pm 0.06$  &  1.33  &  23.8  &  0.56  &  44.2 \\
86.754 &  H$^{13}$CO$^+$ & {\it J}=1--0 & $0.32\pm 0.07$  &  1.50  &  24.8  &  0.65  &  40.1 \\
86.847  &  SiO & {\it J}=2--1 & $2.26\pm 0.08$  &  1.19  &  19.6  &  0.66  &  42.7 \\
88.632  &  HCN & {\it J}=1--0  & $7.58\pm 0.09$  &  1.58  &  26.0  &  0.41  &  44.0 \\
89.189  &  HCO$^+$ & {\it J}=1--0 & $2.48\pm 0.25$  &  1.49  &  20.1  &  0.61  &  46.0 \\
90.664  &  HNC & {\it J}=1--0  & $2.19\pm 0.09$  &  1.58  &  26.3  &  0.48  &  44.4 \\
90.979  &  HC$_3$N & {\it J}=10--9 &  $2.17\pm 0.09$  &  1.10  &  20.8  &  0.67  &  43.6 \\
93.174  &  N$_2$H$^+$ & {\it J}=1--0 & $2.53\pm 0.09$  &  1.31  &  23.7  &  0.63  &   45.5 \\
94.405 &  $^{13}$CH$_{3}$OH & $2_{-1}$--$1_{-1}${\it E}  & $0.28\pm 0.07$  &  1.45  & 27.3  &  0.44  &  46.2 \\
95.169	 &  CH$_{3}$OH & $8_{00}$--$7_{10}${\it A} & $1.56\pm  0.09$  &  0.73  &  15.0  &  0.59  &  44.3 \\
95.914 &  CH$_{3}$OH & $2_{10}$--$1_{10}${\it A} & $0.45\pm 0.11$  &  1.56  &  28.2  &  0.23  &  $-47.1$ \\
96.413 &  C$^{34}$S & {\it J}=2--1 & $0.50\pm 0.07$  &  1.23  &  24.2  &  0.26  &  $-40.6$ \\
96.741 &  CH$_{3}$OH & $2_{00}$--$1_{00}${\it A} & $5.79\pm 0.09$  &  1.36  &  25.4  &  0.57  &  44.7 \\
97.981  & CS & {\it J}=2--1 &  $5.42\pm 0.06$  &  1.54  &  28.7  &  0.40  &  44.1 \\
99.300  &  SO &  $3_2$--$2_1$ & $1.83\pm 0.08$  &  1.31  &  23.9  &  0.56  &  43.8 \\
100.076  &  HC$_3$N & {\it J}=11--10 & $2.20\pm 0.12$  &  0.94  &  19.6  &  0.71  &  43.3 \\
101.478   &  H$_2$CS & $3_{13}$--$2_{12}$ & $0.30\pm 0.11$  & \nodata  & \nodata  &  \nodata  &  \nodata \\
102.064  &  NH$_2$CHO &  $5_{14}$--$4_{14}$	& $0.04\pm 0.14$  & \nodata  & \nodata  &  \nodata  &  \nodata \\
103.040  &  H$_2$CS & $3_{03}$--$2_{02}$ & $0.18\pm 0.12$  &  \nodata  & \nodata  &  \nodata  &  \nodata
\enddata
\tablenotetext{a}{Refers to the $T_{\rm MB}$ at $(l, b, \VLSR)\!=\!(-0\dotdeg 40, -0\dotdeg22, -80\,\kms)$.}
\tablenotetext{b}{The size parameter is calculated as $S\!=\!D {\rm tan}\left( \sqrt{\sigma_{l}\sigma_{b}} \right) $ (Solomon et al. 1987), where $D$ is the distance to the cloud ($=\!8.3$ kpc), and $\sigma_{l}$ and $\sigma_{b}$ are the dispersions along the Galactic longitude and latitude, respectively.  Only the data with $T_{\rm MB}\!\geq\!3\sigma$ are used to calculate the dispersions ($\sigma_{\rm V}$ as well).}
\end{deluxetable}

\section{Results}
Eighteen of the 21 lines were detected from the center of CO--0.40--0.22 (Table 1).  We calculated the size and velocity dispersion of the cloud using the data cube of the detected lines.  We used the distance to the cloud $D\!=\!8.3$ kpc.  All the detected lines show that CO--0.40--0.22 has a compact appearance ($S\!<\!1.6$ pc).  The broad velocity width ($\sigma_{\rm V}\!>\!20$ \kms ) is also common, except for the CH$_{3}$OH $8_{00}$--$7_{10}${\it A} line, which has a high upper-state energy ($E_{\rm u}/k\!\simeq\!84$ K). The HC$_3$N {\it J} = 11--10 line also shows a narrower velocity width and more compact appearance compared to the lower-{\it J} transitions of the same molecule.  In Table 1, we also included the ellipticity ($e$) and the position angle (P.A.) with respect to the Galactic plane.  Except for three low intensity lines (c-C$_3$H$_2$, CH$_{3}$OH, $2_{10}$--$1_{10}${\it A}, and C$^{34}$S), the P.A. of the cloud is concentrated around 45\arcdeg, while the ellipticity ranges from 0.40 to 0.71.    

We present the velocity-integrated map and a position--velocity map of the SiO {\it J} = 2--1 line, as it represents the spatial velocity structure of CO--0.40--0.22 (Fig. 2).  CO--0.40--0.22 has an oval shape with the major axis aligned with the shell 2.  It  consists of an intense component with a shallow velocity gradient and a less intense high-velocity wing.  Shock probe lines, such as the SiO, SO, CH$_3$OH, and HCN {\it J} = 4--3 lines, commonly show this behavior.  Neither an expanding feature nor a cavity appears in CO--0.40--0.22.

A gas mass of $M_{\rm gas}\!=\!10^{3.6}\,M_{\sun}$ is derived from the HCN {\it J} = 1--0 line intensity using the large velocity gradient model.  Here, we assumed the kinetic temperatre to be $T_{\rm k}\!=\!60$ K which is derived for another well-studied HVCC CO 0.02--0.02 (Oka et al. 1999), and the [HCN]/[H$_2$] abundance ratio of $10^{-7.3}$ which is derived for the Sgr B1 cloud (Tanaka et al. 2009; Oka et al. 2011).  The density was chosen to be $n({\rm H}_2)\!\geq\!10^{6.5}$ cm$^{-3}$ to reproduce the observed HCN {\it J}=1--0 and {\it J}=4--3 intensities at $(l ,b, \VLSR)\!=\!(-0.40, 0.22, -80\,\kms)$, $T_{\rm MB}^{1-0}\!=\!7.6$ K and $T_{\rm MB}^{4-3}\!=\!16.3$ K.  A size parameter of 1.0 pc and velocity dispersion of 20 \kms\ give a virial theorem mass of $M_{\rm VT}\!=\!1\!\times\! 10^6\,M_{\sun}$.   This yields a very large virial parameter, $M_{\rm VT}/M_{\rm gas}\!\sim\!300$, indicating that the gas mass is definitely insufficient to bind the cloud by its self-gravity.  The kinetic energy amounts to $E_{\rm kin}\!=\!10^{49.7}$ erg if the velocity dispersion is dominated by random motion.  If it is expanding at $V_{\rm exp}\!=\!40$ \kms\ (half of the velocity extent), the kinetic energy becomes $E_{\rm kin}\!=\!10^{49.8}$ erg.  CO--0.40--0.22 is characterized by a rather featureless spatial velocity structure.   This is in sharp contrast with CO 0.02--0.02 (Oka et al. 1999; 2008) or CO 1.27+0.01 (Oka et al. 2001; Tanaka et al. 2007), which are HVCCs containing expanding shells or emission cavities.

\begin{figure}[htbp]
\begin{center}
\includegraphics[width=0.4\textwidth]{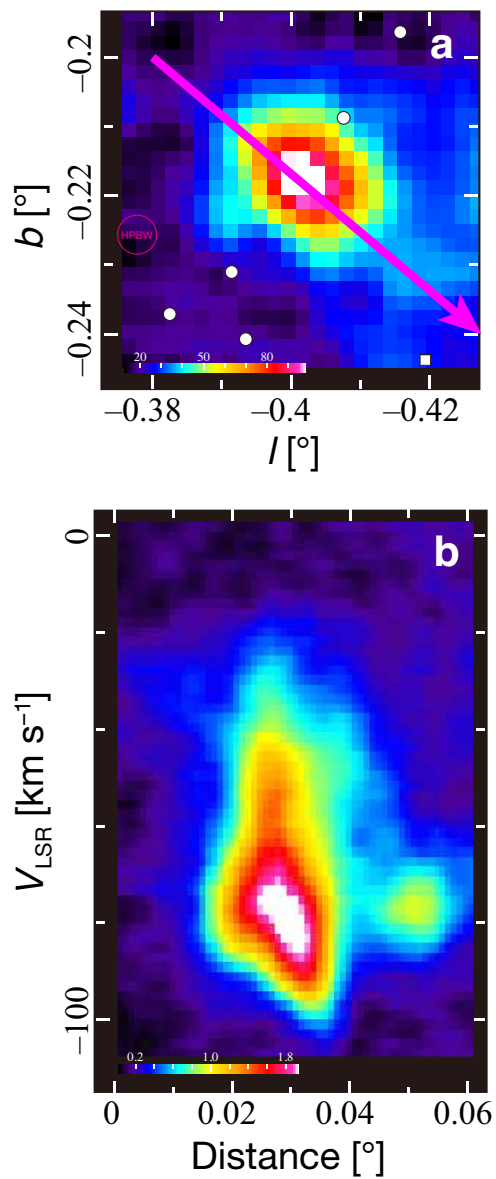}
\end{center}
\caption{(a) Map of SiO {\it J} = 2--1 emission integrated over velocities from $-110$ to $0$ \kms.  Color represents the integrated intensity in units of K \kms.  A magenta circle shows the half-power beamwidth (HPBW) of the telescope at 86 GHz.  White filled circles denote the positions of X-ray sources (Muno et al. 2006).  White filled square shows the position of IRAS 17423--2924.  (b) Position--velocity map of SiO {\it J} = 2--1 emission along the magenta arrow indicated in panel {\bf a}.  Color represents the main-beam temperature in units of K.  The horizontal axis indicates the angular distance from $(l, b)\!=\!(-0.38\arcdeg , -0.20\arcdeg)$ along the arrow. }
\label{fig2}
\end{figure}

\section{Discussion}
\subsection{Origin of CO--0.40--0.22}
We inspected the AKARI point source catalog (Yamauchi et al. 2011) and the Chandra X-ray source catalog (Muno et al. 2006) to search for counterparts to CO--0.40--0.22.  Five X-ray sources are found in the current field of view (FOV), whereas no AKARI point source was found (Fig. 2a).  A far-IR point source, IRAS 17423--2924, resides in the lower right edge of the FOV.  No X-ray  source or IR source overlaps with the CO--0.40--0.22 main body in the plane of the sky.

Tanaka et al. (2014) presented four interpretations for the broad velocity width nature of HVCCs: acceleration by an external expanding shell, bipolar outflow, expansion driven by supernovae (SNe), and rotation around a dark massive object.  The lack of counterparts of CO--0.40--0.22 at other wavelengths disfavors formation scenarios based on internal explosive events.  Further, the absence of an expanding feature or a cavity also reduces support for internal explosion/outflow scenarios.  The associated expanding shell (shell 2) could have distributed kinetic energy to CO--0.40--0.22.  The kinetic energy of the shell 2, was derived to be $\sim 7\!\times\!10^{50}$ erg.  Because CO--0.40--0.22 occupies only $1/400$ of the shell 2 in solid angle, the energy share is only $\sim\! 10^{48.2}$ erg.  This is 1.5 orders of magnitude smaller than the kinetic energy of CO--0.40--0.22, $E_{\rm kin}\!=\!10^{49.7}$ erg.  

Another possibility is that the plunging of a small cloud into the CMZ.  This could explain the broad velocity width of CO--0.40--0.22, and such a high-velocity object could belong to the Galactic halo.  However, such a plunge should make cometary srtucture.  The featureless appearance of CO--0.40--0.22 renders the plunging scenario implausible.  

\subsection{Gravitational Kick Model}
A promising candidate for an explanation of the formation scenario involves a ``gravitational kick'' to a small incoming cloud by a compact source.  A point-like mass of $10^5$ $M_{\sun}$ can accelerate a cloud coming from infinity to $\sim\!90$ \kms\ at a distance of $0.1$ pc.  This scenario can easily explain the compact appearance and very large velocity width of CO--0.40--0.22.  However, we must then accept the presence of a massive compact ($<\!0.1$ pc) object.  The association with the expanding shell supports this scenario because it can provide the massive object with a number of small clouds.  For a semiquantitative comparison, we simulated the position--velocity behavior of an incoming cloud.

We placed a cloud of 200 test particles with a $\sigma\!=\!0.2$ pc Gaussian centered at $(\Delta X, \Delta Y)\!=\!(10\,\mbox{pc}, Y_0)$ offset from the massive object. The initial velocity was set to $(v_{\rm X}, v_{\rm Y})\!=\!(-10\,\kms , 0\,\kms )$.  In this simulation, the free parameters are the central mass ($M$), initial $Y$ value ($Y_0$), line-of-sight angle ($\phi$), and  elapsed time ($t$).  The parameter ranges we searched are, $M\!=\!10^3\mbox{--}10^6$ $M_{\sun}$, $Y_0\!=\!(-3.0)\mbox{--}(-0.5)$ pc, $\phi\!=\!0\arcdeg\mbox{--}90\arcdeg$, and $t\!=\!0\mbox{--}10^6$ yrs.  After a number of trials, we found that the parameter set $M\!=\!10^5$ $M_{\sun}$, $y_0\!=\!(-1.8)\mbox{--}(-1.0)$ pc, $\phi\!=\!45\arcdeg$, and $t\!\sim\!7\!\times\!10^5$ years reproduces the position--velocity behavior of CO--0.40--0.22 well.  Fig. 3 shows the time evolution of two initially Gaussian clouds of $\sigma\!=\!0.2$ pc with $Y_0\!=\!(-1.0)$ pc (cloud 1) and $Y_0\!=\!(-1.8)$ pc (cloud 2).  The locus of the gravitational source corresponds to the center of CO--0.40--0.22, from which the positive-side highest velocity emission arises.

\begin{figure}[htbp]
\begin{center}
\includegraphics[width=0.5\textwidth]{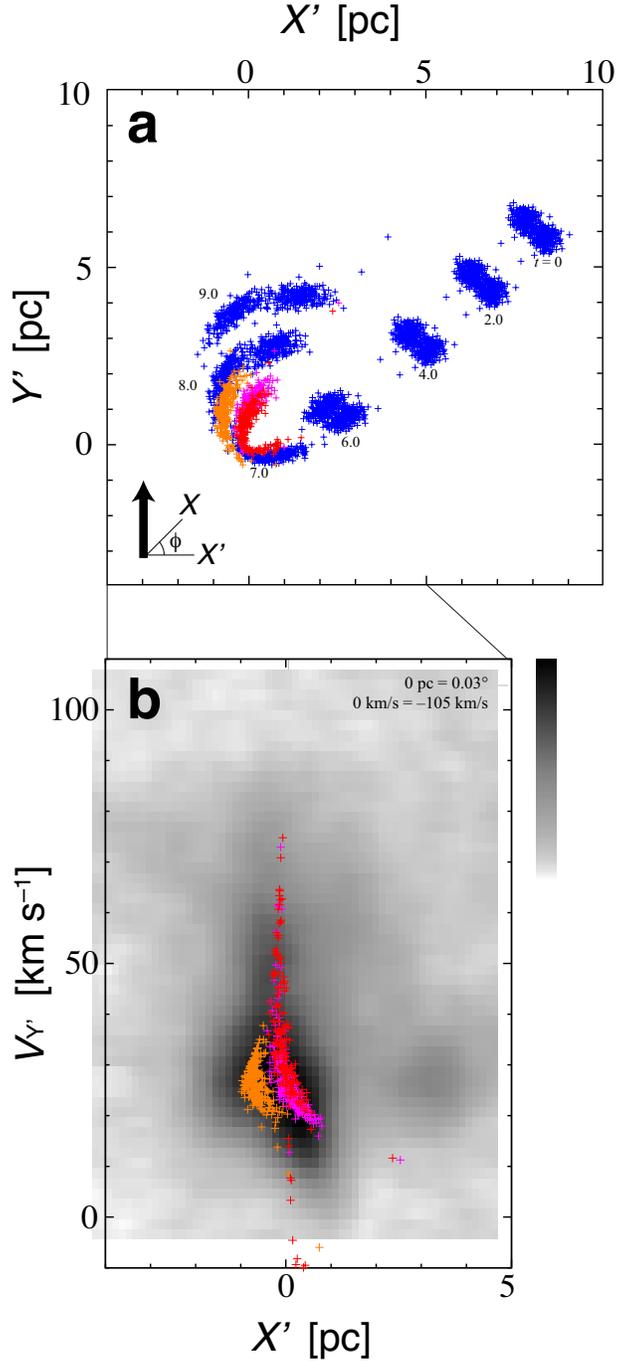}
\end{center}
\caption{(a) Time evolution of clouds 1 and 2 (see text) in the orbital plane.  $X$--$Y$ coordinates are rotated by $\phi=45\arcdeg$.  Thick arrow indicates the line-of-sight direction.  (b) Position--velocity plot of the simulated clouds superposed on the SiO {\it J} = 2--1 map (gray scale).  The vertical axis shows the radial velocity, and the horizontal axis shows the transverse offset from the mass point.  The elapsed times are $t\!=\!7.0\!\times\!10^5\,\mbox{yr}$ (red) and $7.2\!\times\!10^5\,\mbox{yr}$ (magenta) for cloud 1, and $t\!=\!7.6\!\times\!10^5\,\mbox{yr}$ for cloud 2 (orange). }
\label{fig3}
\end{figure}

What is this massive compact object?  Supposing the gravitational kick scenario is valid, the object must be as large as $10^5$ $M_{\sun}$ with a radius that is significantly smaller than 0.1 pc (pericenter distance).  These mass and size values correspond to an average mass density of $\rho\!\simeq\!2\!\times\! 10^7$ $M_{\sun}\, \mbox{pc}^{-3}$.  This mass density is comparable to that of the core of M15, which is one of the most densely packed (core-collapsed) globular clusters in the Milky Way Galaxy (Djorgovski \&\ King 1984).  However, the total mass within 0.05 pc of the center of M15 is only 3400 $M_{\sun}$ (van den Bosch et al. 2006); thus, the generation of CO--0.40--0.22 by a globular cluster seems implausible.  The lack of counterparts at other wavelengths is inconsistent with the massive stellar cluster interpretation, unless the cluster consists almost entirely of dark stellar remnants, such as neutron stars and BHs.  Therefore, it is most likely that the massive compact object responsible for the formation of CO--0.40--0.22 is an invisible point-like mass of $\sim\! 10^5$ $M_{\sun}$, i.e., an IMBH.  

The chance probability of a BH-molecular cloud encounter can be estimated by the gas volume filling factor.  The volume filling factor of dense [$n({\rm H}_2)\!\geq\!10^4$ cm$^{-3}$] molecular gas was estimated to be $>\! 0.1$ (Morris and Serabyn 1996).  We expect the larger volume filling factor for the less dense ($<\!10^4$ cm$^{-3}$) molecular gas and the gas compression due to the encounter with a BH.  Thus severall IMBH in the CMZ are sufficient to account for the detection of such an object.

\subsection{Origin of the IMBH}
The origin of such a ``massive'' IMBH is controversial.  Theoretical studies state that a $10^3$ $M_{\sun}$ IMBH can be formed at the center of a dense stellar cluster of several times $\,\!10^5$ $M_{\sun}$ through the runaway merging of massive stars (Marchant \&\ Shapiro 1980).  The presence of such IMBHs is suggested by the features of several globular clusters in the Milky Way Galaxy (Gebhardt et al. 2002; Gerssen et al. 2002; Noyola et al. 2008), although their suggestions are received with great caution (e.g., Strader et al. 2012).  Such dense stellar clusters with IMBHs sink to the galactic center by dynamical friction and then merge to form a more massive BH (Ebisuzaki et al. 2001).  However, CO--0.40--0.22 is located at least $60$ pc from the Galactic nucleus.  The relation between the BH mass and the stellar system mass (e.g., Kormendy \&\ Ho 2013) indicates that a $10^5$ $M_{\sun}$ BH may be involved in a stellar system of $\sim\!10^8$ $M_{\sun}$, which falls into the mass range of dwarf galaxies.  Recently, SMBHs have been found at the centers of dwarf galaxies (Reines et al. 2011; Seth et al. 2014) and, in the vicinity of the Milky Way Galaxy, over 20 dwarf satellite galaxies have been discovered to date.  It is believed that large galaxies such as the Milky Way have grown to their present form by cannibalizing their smaller neighbors.  Thus, it is natural to suggest that the $10^5$ $M_{\sun}$ BH in the Galactic CMZ was the nucleus of a cannibalized dwarf galaxy.


Although the IMBH model is speculative, the case of CO--0.40--0.22 may open a new perspective on the search for BH candidates.  The total number of stellar-mass BHs in the Milky Way Galaxy is estimated to be more than $10^8$ (Agol \&\ Kamionkowski 2002).  If one of these BHs encounters a molecular cloud, it leaves a kinematical signature in the cloud, whether the BH is luminous or not.  The search for compact high-velocity features using molecular lines is an effective method of seeking stellar-mass BHs in the Galactic disk.  For example, such an ultrahigh-velocity molecular feature has been detected in the molecular cloud adjacent to the supernova remnant W44 (Sashida et al. 2013).  Further, some of the HVCCs in the Galactic CMZ (Oka et al. 1998; 1999; 2012; Tanaka et al. 2007; 2014) also provide possible candidates for nonluminous BHs.  High-resolution aperture synthesis imaging of such HVCCs will reveal their detailed spatial structures and kinematics, which are crucial to confirming the presence of BHs within them.  Further, the detection of such compact high-velocity features in nearby galaxies will increase the number of IMBH candidates and thereby generalize our gravitational kick interpretation.  This new perspective will greatly contribute to our understanding of galactic evolution.

\acknowledgments
This paper is based on observations conducted using the Nobeyama Radio Observatory (NRO) 45-m telescope and ASTE.  The NRO is a branch of the National Astronomical Observatory of Japan (NAOJ), National Institutes of Natural Sciences.  We are grateful to the NRO staff and all the members of the ASTE team for operation of the telescope.  Observations with ASTE were conducted remotely from NRO using NTT's GEMnet2 and its partner research and education (R\&E) networks, which are based on the AccessNova collaboration between the University of Chile, NTT Laboratories, and the NAOJ.  We also thank F. Yusef-Zadeh for suggesting the gravitational kick scenario.  T.O. acknowledges support from JSPS Grant-in-Aid for Scientific Research (C) No. 24540236.

\end{document}